\documentclass{article}
\usepackage{amssymb}

\input{tcilatex}

\begin{document}

\begin{center}
\textbf{Decoherence Rates in Large Scale Quantum Computers and Macroscopic
Quantum Systems}\bigskip

B J DALTON\bigskip

\textit{Australian Research Council Centre of Excellence for Quantum-Atom
Optics}\\[0pt]
\textit{and}\\[0pt]
\textit{Centre for Atom Optics and Ultrafast Spectroscopy, Swinburne
University of Technology, Melbourne, Victoria 3122, Australia}\bigskip
\end{center}

\textbf{Abstract}. Markovian regime decoherence effects in quantum computers
are studied in terms of the fidelity for the situation where the number of
qubits $N$\ becomes large. A general expression giving the decoherence time
scale in terms of Markovian relaxation elements and expectation values of
products of system fluctuation operators is obtained, which could also be
applied to study decoherence in other macroscopic systems such as Bose
condensates and superconductors. A standard circuit model quantum computer
involving three-state lambda system ionic qubits is considered, with qubits
localised around well-separated positions via trapping potentials. The
centre of mass vibrations of the qubits act as a reservoir. Coherent one and
two qubit gating processes are controlled by time dependent localised
classical electromagnetic fields that address specific qubits, the two qubit
gating processes being facilitated by a cavity mode ancilla, which permits
state interchange between qubits. With a suitable choice of parameters, it
is found that the decoherence time can be made essentially independent of $N$%
.{\small \medskip }

\section{\textbf{Introduction}{\protect\small \protect\medskip }}

The topic of quantum computation has developed enormously since the
theoretical work of Feynman \cite{Feynman80a} and Deutsch \cite{Deutsch85a}
in the 1980s. Along with topics such as quantum teleportation, quantum
cryptography and quantum measurement theory, quantum computation is one part
of the expanding field of quantum information science \cite{Nielsen00a}.
Much of the current interest has been stimulated by potential applications
of quantum computers in situations where they are expected to out-perform
classical computers, such as in quantum algorithms for factoring large
numbers \cite{Shor94a} and searching large data bases \cite{Grover97a}, or
in simulating the behavior of quantum systems \cite{Feynman80a}. The
implementation of quantum computers has mainly focused on the standard
quantum circuit model. Here quantum information is stored as entangled
states of an array of two-state systems (qubits) which are initially
prepared with all qubits in one state, the algorithm is then implemented as
a sequence of unitary operations involving one or two qubits at a time, and
the result for the computation is provided by a measurement on the final
state of the quantum computer. Other approaches have also been proposed,
including adiabatic quantum computation \cite{Fahri01a}, continuous
variable, topological or holonomic quantum computation \cite{Gottesman01a}, 
\cite{Freedman03a}, \cite{Duan01a} and quantum computation based on a
sequence of projective measurements instead of unitary quantum gates \cite%
{Knill01a}, \cite{Nielsen99a} - including those using cluster states \cite%
{Raussendorf01a}). A variety of physical realisations of the quantum circuit
model have been proposed, differing in the nature of the qubit system (e.g ,
hyperfine ground states and metastable excited states in ions and neutral
atoms, one photon states in optical systems with two polarisation modes,
...), the way the qubits are prepared in the initial state (e. g., optical
pumping - ionic and neutral atom qubits), the method used for the gating
process (e. g., two photon-resonant laser fields - one qubit gates in ions
and atoms, one photon resonant laser fields combined with photons in high Q
optical cavities - two qubit gates for atoms in cavity QED based systems,
cold collisions between qubits - two qubit gates for neutral atoms, ..) and
the technique used to detect qubit states (e. g., quantum jump techniques
for ionic qubits). Small scale demonstrations of quantum computation have
already been achieved, such as implementing the Deutsch-Josza algorithm \cite%
{Gulde03a} and factoring small numbers \cite{Vandersypen01a}. Quantum
computer architectures, which specify the components and how they are
integrated within a coherent plan, provide useful guides for the long term
research program needed to develop practical quantum computers, and allow
for developments of the basic architecture to overcome problems revealed as
the research program is implemented. For example, ion trap quantum computers
have significantly evolved from the original Cirac-Zoller single trap
proposal \cite{Cirac95a} to an architecture involving qubits shuttling
between a memory region and a processor region in order to build a larger
scale system \cite{Kielpinski02a}. The implementation of quantum computers
via quantum optical systems (atoms, ions and photons) is one of the more
promising routes to follow \cite{Monroe02a}, \cite{Cirac04a} and the present
paper deals with such a system. A set of criteria for the successful
realisation of circuit model quantum computers has been formulated by
DiVincenzo \cite{DiVincenzo00a}, and comprehensive surveys of current
knowledge of the subject are contained in recent reviews \cite{LANL04a} and
textbooks \cite{Nielsen00a}.\medskip

The idealised unitary evolution of the circuit model quantum computer does
not occur in reality because the quantum computer interacts with the
environment, both during gating processes and when no gating is taking
place. The loss of unitarity is referred to as decoherence, and can be
quantified in terms of the fidelity. This specifies how close the actual
behavior of the qubit system density operator is to its idealised behavior
when only coherent gating processes take place and system-reservoir
interactions are switched off. The time for the fidelity to change
significantly from unity defines the decoherence time scale. Other measures
of decoherence time scales have also been proposed \cite{Fedichkin03a}.
Decoherence is also of interest in other macroscopic systems such as Bose
condensates and superconductors, and its general effect in macroscopic
systems is important in quantum measurement theory and to understanding how
classical behavior emerges \cite{Zurek03a}, \cite{Guilini96a}. In the case
of quantum computers, one of the DiVincenzo criteria for effective quantum
computation is that decoherence time scales must be much larger than gating
time scales and the time intervals when gating processes are absent, and
hence a determination of the decoherence time scale is important in
evaluating any specific quantum computer architecture. Decoherence is the
enemy of quantum computation and a number of methods have been proposed for
combating its effects. These include active (or error correcting) methods -
quantum error correction \cite{Shor95a}, \cite{Calderbank96a}, \cite%
{Steane96a}, dynamical decoupling techniques \cite{Viola98a}, and passive
(or error avoiding) methods - decoherence-free subspaces \cite{Zanardi97a}, 
\cite{Lidar98a}, \cite{Duan98a}, \cite{Beige00a}, topological or holonomic
quantum computing \cite{Gottesman01a}, \cite{Freedman03a}, \cite{Duan01a}.
Combined methods, such as decoherence-free subspaces plus dynamical
decoupling have also been proposed \cite{Byrd03a}. Quantum error correction
implies a large overhead of redundant qubits and dynamical decoupling
involves control pulses with time scales much shorter than reservoir
correlation times, so in general terms it is desirable to implement error
avoiding methods and use error correcting \ as a back up. It is well known
that if the loss of fidelity in each gating process is kept below a certain
threshold (estimated as being between $10^{-3}$ and $10^{-4}$- see \cite%
{Preskill99a}, \cite{LANL04a}) then fault-tolerant quantum computing is
possible using error correcting codes \cite{Preskill98a}, \cite{Knill98a}.
Error avoiding methods will be the focus of the present paper, with the aim
of keeping the fidelity loss below such a threshold. The scalability of a
quantum computer architecture is another important DiVincenzo criterion. The
large qubit case is important for implementing quantum computers in
situations where they are expected to out-perform classical computers, such
as factoring large numbers and searching large data bases \cite{Nielsen00a}.
In this paper, the primary aim will be to study decoherence effects in
quantum computers for the situation where the number of qubits $N$\ becomes
large. \medskip

In general \cite{Namiki03a}, the temporal behavior of a quantum system
coupled to a zero temperature reservoir is of three distinct types,
depending on the time regime: (a)\ Quadratic behavior at short times ($t\ll
\tau _{c}$) (b) \ Exponential decay at intermediate times ($\tau _{d}\approx
t\gg \tau _{c}$) and (c) Power law behavior at long times ($t\gg \tau _{d}$%
). Here $\tau _{c}$\ is the reservoir correlation time, $\tau _{d}$\ the
system decay time. At non-zero temperature $T$, a thermal time scale $\tau
_{b\,}=\hbar /k_{B}T$\ is also involved. The short time regime is associated
with the Quantum Zeno effect, the intermediate time regime involves
Markovian decay and the long time regime behavior is due to a lower bound in
the energy spectrum.\medskip

For quantum computers, the physical relevance of the short time regime is
not clear, since (apart from architectures making use of dynamical
decoupling methods for error correction purposes \cite{Viola98a}) most
feasible measurements and gating processes are likely to require time scales
much longer than the reservoir correlation time. As Markovian theories for
the intermediate time regime indicate that decoherence times decrease
rapidly as the number of qubits $N$\ increases, the decoherence time may
finally become comparable to the reservoir correlation time. Apart from
creating consistency problems for the theory, this could place a limit on
quantum computer size. The relationship between the short and intermediate
time regimes has been studied for a simple case of $N$\ two-state systems,
all initially in the lower state and coupled to the electromagnetic (EM)
field in the vacuum state. A perturbation treatment correct to second order
in the coupling constants \cite{Dalton04a} shows that the fidelity initially
decreases quadratically for times less than $\tau _{c}$, then reaches a
minimum and eventually returns to unity. The Markoff theory fidelity equals
one at all times. The effect is due to the non RWA terms in the
system-reservoir coupling, and may merely reflect the artificial nature of
the uncorrelated system-reservoir initial state. Alternative treatments of
the short time regime based on methods preserving unitarity (\cite%
{Tolkunov04a}, \cite{Braun01a}) may overcome these difficulties for studying
systems with very large $N$. \ A preliminary study \cite{Dalton03a} on a
standard quantum computer model with two-state qubits obtained a quadratic
behavior of the fidelity in the short time non-Markovian regime.\medskip

The present work deals with the intermediate time regime. The intermediate
time regime is more physically relevant for reasons discussed above (the
quantum computer model we study does not involve the use of dynamical
decoupling), and here Markoff theory can be used. For internal consistency,
the decoherence time scale $\tau _{D}$\ must be long compared to $\tau _{c}$%
. Markovian expressions for the intermediate time behavior of the rate of
change of fidelity are obtained for the general case where the qubit system
(including any ancilla) are in a pure state and the reservoir (which may
involve several components) is in a thermal state. The initial rate of
change of the fidelity defines the decoherence rate and its inverse is the
decoherence time scale $\tau _{D}$. This result gives the decoherence time
scale in terms of Markovian relaxation elements and expectation values of
products of fluctuation operators for the decohering quantum system. The
expression is quite general and may have applications for decoherence in
other macroscopic systems, such as Bose condensates or superconductors. For
the quantum computer model studied, the characteristic decoherence time
scale is evaluated at finite temperature for specific qubit states (such as
Hadamard and GHZ states) in the situation of no gating processes occuring
(memory decoherence). Decoherence effects on quantum computers due to one
and two qubit gating processes are studied for zero temperature (gating
decoherence). The zero temperature case should be most favourable for long
decoherence times, and short decoherence times even at zero temperature
during gating processes would be ominous for implementing large scale
quantum computers.\medskip

As mentioned previously, architectures involving ionic qubits are amongst
the most promising for possible implementations of circuit model quantum
computers \cite{Monroe02a}, \cite{Cirac04a}, \cite{LANL04a}. However,
directly scaling up the original Cirac-Zoller model \cite{Cirac95a} to large
numbers of qubits is difficult not only because it is hard to trap large
numbers of ions in a single trap, but also because using a collective
vibrational mode as an ancilla for two qubit gates becomes impossible when
the mode density becomes very large. An extensive study of decoherence
effects as qubit numbers are increased in Cirac-Zoller type quantum
computers where a vibrational mode is used for gating purposes, has been
carried out by Plenio and Knight \cite{Plenio96a}, \cite{Plenio97a} for both
two state and three state lambda system qubits, and applied to real ions.
Limitations on the size of quantum computers (and therefore on the size of
numbers factorisable via the Shor algorithm \cite{Shor94a}) based on this
model were found. Even allowing for error correction this limit was quite
small. A solution to the problem of scaling up qubit numbers is being
developed using shuttled qubits \cite{Kielpinski02a}, and a different
approach involving a 2D array of trapped ions with a control qubit moved
above its target target qubit to carry out a two qubit gate via a
collisional process has also been suggested \cite{Cirac00a}, \cite%
{Calarco01a}. The 2D arrays might be based on elliptical rather than linear
ion traps \cite{DeVoe98a}. Another approach not involving ion shuttling has
also been proposed, with two qubit gating via Raman laser pulses applied to
the pair of qubits \cite{Duan04a}. Architectures involving high Q cavities
are also attractive \cite{Monroe02a}, \cite{Cirac04a}, \cite{LANL04a}, and
it is worthwhile to try to combine these with ionic qubits. Cavity QED
systems involve using an optical (or microwave) cavity mode photon as an
ancilla to enable two qubit gating to occur for atomic or ionic qubits, and
were in fact amongst the earliest proposals for quantum computers \cite%
{Pellizzari95a}, \cite{Domokos95a}, \cite{Turchette95a}. High-finesse
optical cavities have now been developed \cite{Sauer04a} which allow a large
number of atomic qubits to be enclosed under strong coupling conditions, so
that coupling of the cavity mode to the qubit system is faster than both
cavity photon loss and qubit decay via spontaneous emission. In such
circumstances, two qubit gating via cavity photons can occur much faster
than decoherence due to the loss processes. However, it is difficult to
create strong coupling conditions for ionic qubits, since the small optical
cavity volume can interfere with the ion trap and the presence of electrodes
can interfere with the cavity mode \cite{Duan04b}, though single ion
trapping inside a cavity under weak coupling regime conditions has been
realised \cite{Guthohrlein01a}, \cite{Mundt02a}. There have been several
proposals for scalable quantum computers involving ionic qubits in optical
cavities under weak coupling conditions \cite{Beige00a}, \cite{Beige00b}, 
\cite{Pachos02a}, \cite{Tregenna02a}, \cite{Duan04b}, involving
probabilistic entanglement protocols or using dissipation to confine the
evolution in decoherence-free subspaces. The optical cavity may be arranged
with its axis perpendicular to the array of ions, so that two qubits at a
time are in the cavity. Strong coupling regime proposals have also been
formulated \cite{Beige04a}. In the present paper, we consider a combined
ionic qubit and high Q cavity quantum computer architecture with a large
number of qubits in the cavity. Our model is somewhat similar to the two
qubit case considered by Tregenna, Beige and Knight \cite{Tregenna02a}, but
now we consider a large number of qubits and also allow for their
vibrational motion rather than treat them as stationary. As the emphasis of
the present work is to examine the effects of decoherence on the system
considered by Tregenna et al \ \cite{Tregenna02a} as qubit numbers are
increased, the same parameters as in their work will be used, rather than
those for real ions. The issue of developing a theory for real ions is
discussed in the last section of the paper. Memory decoherence for the
Cirac-Zoller model \cite{Cirac95a} due to the effects of vibrational motion
has also been studied by Garg \cite{Garg96a}. Note that in the present case
(unlike in the work of Plenio and Knight \cite{Plenio96a}, \cite{Plenio97a}%
), the vibrational modes act as a reservoir rather than as an ancilla to
facilitate gating processes. \medskip 

A standard model involving $N$\ ionic qubits is considered, the overall
architecture being illustrated in figure 1. Each qubit is in a three-state
lambda system \cite{Beige00a}, rather than a two-state system as previously
treated in the short time regime \cite{Dalton03a}. The qubit states are the
two lower states 0, 1. The quantum computer system also includes a high Q
cavity mode, which is coupled to the qubits and acts as an ancilla. Lambda
systems, as well as facilitating Raman gating processes, should result in
qubits that are less vulnerable to spontaneous emission (SE) based
decoherence, the upper state 2 only being occupied during gating processes.
Reducing SE by having qubits located in a high Q cavity is also desirable,
and the cavity mode also facilitates two qubit processes. Our model involves 
$N$\ ionic qubits localised around well-separated positions via trapping
potentials, and the centre of mass vibrational motions of the qubits are now
treated. Coherent one and two qubit gating processes are controlled by time
dependent localised classical EM fields that address specific qubits. The
one qubit gating process involves weak two-photon resonant Raman gating
fields, well detuned from one photon resonance. The two qubit gating
processes are facilitated by the cavity mode ancilla, which permits state
interchange between qubits. For the two qubit gating process, resonant
gating fields coupled to the 2-1 transition for the control qubit, and
coupled to the 2-0 transition for the target qubit. The cavity mode is
resonant with the 2-1 transition, but uncoupled to the well-detuned 2-0
transition (as in \cite{Tregenna02a}). Two qubit gating processes take place
in decoherence-free subspaces \cite{Beige00a}, \cite{Tregenna02a}. In our
model, the reservoir (or bath) has three constituents. The three-state
qubits are coupled to a bath of spontaneous emission modes, and the cavity
mode is coupled to a bath of cavity decay modes. For large $N$\ the numerous
vibrational modes of the ionic qubits also act as a reservoir, with
Lamb-Dicke coupling to the internal qubit coordinates, cavity mode and
gating fields. Non-RWA couplings are included. All qubit interaction terms
(electric dipole, Rontgen, diamagnetic, ionic current) with these three
baths are examined and the important contributions to the decoherence rate
found. The qubit-bath coupling is amplitude coupling via $\sigma _{\pm
}^{ia} $\ optical coherence operators. In addition to these fundamental
causes of decoherence, technical shortcomings in the implementation of the
computer model can also cause decoherence. For example, the trapping
potential producing the array of qubit trap sites could be subject to
fluctuations. Such trapping potentials could be provided by off resonant
near-classical optical fields or by magnetic fields, and these could
fluctuate. Decoherence effects due to fluctuations in these fields could
have significant effects \cite{Schmidt-Kaler03a} and should be evaluated .
However, we will concentrate in this paper on fundamental causes of
decoherence, especially the effects of qubit vibrations, and technical
causes will be left to a later time.\medskip

The plan of this paper is as follows. In section 2 we set out the
Hamiltonian for the quantum computer model and derive our general expression
for the decoherence time scale by applying Markovian evolution theory. In
section 3 the decoherence time scales are evaluated for the no-gating, one
qubit gating and two qubit gating cases. A summary of the main results of
the paper is presented in section 4 along with a discussion of extensions of
the theory for real ions.\bigskip

\section{Theory{\protect\small \protect\medskip }}

\subsection{Hamiltonian{\protect\small \protect\medskip }}

The total Hamiltonian for the system can be written as%
\begin{equation}
H=H_{S}+H_{C}+H_{B}+V_{S}+V_{I}
\end{equation}%
where the Hamiltonian for the qubit system and cavity mode ancilla is%
\begin{equation}
H_{S}=\sum_{ia}\hbar \omega _{a}\sigma _{aa}^{i}+\hbar \omega _{b}b^{\dag }b
\end{equation}%
the Hamiltonian for the collective vibrational motions of the qubit system is%
\begin{eqnarray}
H_{C} &=&\frac{1}{2m}\sum_{i\alpha }p_{i\alpha }^{2}+\frac{1}{2}%
\sum_{ij\alpha \beta }V_{ij}^{\alpha \beta }\delta r_{i\alpha }\delta
r_{j\beta }  \label{Eq.HamLattVib} \\
&=&\sum_{K}\hbar \nu _{K}A_{K}^{\dag }A_{K}
\end{eqnarray}%
and the Hamiltonian for the bath of spontaneous emission and cavity decay
modes is%
\begin{equation}
H_{B}=\sum_{k}\hbar \omega _{k}a_{k}^{\dag }a_{k}+\sum_{k}\hbar \xi
_{k}b_{k}^{\dag }b_{k}.
\end{equation}%
\medskip

The coherent coupling Hamiltonian for gating processes in the qubit system is%
\begin{eqnarray}
V_{S} &=&\sum_{i;a=0,1}\hbar (\Omega _{ia}+\Omega _{ia}^{\ast })(\sigma
_{+}^{ia}+\sigma _{-}^{ia})  \nonumber \\
&&+\sum_{i;a=0,1}\hbar (g_{ia}b+g_{ia}^{\ast }b^{\dag })(\sigma
_{+}^{ia}+\sigma _{-}^{ia})
\end{eqnarray}%
The qubits are coupled to both classical fields and the cavity mode. Each
qubit is addressed by localised classical EM fields to facilitate 1 qubit
and 2 qubit gating. For 1 qubit gating $\Omega _{ia}$ for $i$th (gated)
qubit are two photon resonant Raman fields strongly detuned from 0-2 and 1-2
transitions.For 2 qubit gating $\Omega _{i1}$ is resonant with the 1-2
transition for the $i$th (control) qubit and $\Omega _{j0}$ is resonant with
the 0-2 transition for the $j$th (target) qubit \cite{Tregenna02a}. The
cavity frequency $\omega _{b}$\ is resonant with the 1-2 optical transition 
\cite{Tregenna02a}. In the model of Tregenna et al \cite{Tregenna02a} the
cavity mode is only coupled to the 1-2 optical transition $g_{i0}=0$\
\medskip

Finally, the interaction of the qubit system and ancilla with the bath and
centre of mass vibrations is given by 
\begin{eqnarray}
V_{I} &=&\sum_{i;a=0,1}\hbar (\sigma _{+}^{ia}+\sigma
_{-}^{ia})\sum_{k}(g_{k}^{ia}a_{k}+g_{k}^{ia\ast }a_{k}^{\dag })  \nonumber
\\
&&+\hbar b^{\dag }[\sum_{k}(v_{k}b_{k}+w_{k}b_{k}^{\dag
})+\sum_{iK}t_{K}^{i\ast }(A_{K}^{\dag }-A_{K})]+HC  \nonumber \\
&&+\sum_{i;a=0,1}\hbar (\sigma _{+}^{ia}+\sigma
_{-}^{ia})b\sum_{K}p_{K}^{ia}(A_{K}+A_{K}^{\dag })+HC  \nonumber \\
&&+\sum_{i;a=0,1}\hbar (\sigma _{+}^{ia}+\sigma _{-}^{ia})\sum_{K}(\Theta
_{K}^{i}+\Theta _{K}^{i\ast })(A_{K}+A_{K}^{\dag })
\end{eqnarray}%
Terms include electric dipole coupling of qubits to SE modes, quasi-mode
coupling of cavity mode to decay modes, Lamb-Dicke coupling of qubits,
cavity mode, gating field to CM modes. Rontgen and diamagnetic terms are not
included as their effects were shown to be small.\medskip

The centre of mass (CM) displacement of the qubits is related to the
collective vibrational modes of the qubits via%
\begin{eqnarray}
\delta r_{i\alpha } &=&\sum_{K}S_{i\alpha ;K}\sqrt{\frac{\hbar }{2m\nu _{K}}}%
(A_{K}+A_{K}^{\dag }) \\
\sum_{j\beta }V_{ij}^{\alpha \beta }S_{j\beta ;K} &=&m\nu _{K}^{2}S_{i\alpha
;K}  \label{Eq.LatticeVibFreq}
\end{eqnarray}%
\medskip\ where the unitary real matrix $S$\ relates qubit CM displacements $%
\delta r_{i\alpha }$\ $(\alpha =x,y,z)$\ to vibrational normal
coordinates.\medskip

The qubit, cavity mode, centre of mass motion, bath modes operators are

\begin{eqnarray}
\sigma _{+}^{ia} &=&(|2\rangle \langle a|)_{i}\quad \sigma
_{-}^{ia}=(|a\rangle \langle 2|)_{i}\quad a=0,1  \nonumber \\
\sigma _{ab}^{i} &=&(|a\rangle \langle b|)_{i}\quad a\neq b\quad a=0,1 
\nonumber \\
\sigma _{aa}^{i} &=&(|a\rangle \langle a|)_{i}\quad a=0,1,2 \\
\lbrack b,b^{\dag }] &=&1\quad \lbrack a_{k},a_{l}^{\dag }]=\delta _{kl} 
\nonumber \\
\lbrack b_{k},b_{l}^{\dag }] &=&\delta _{kl}\quad \lbrack A_{K},A_{L}^{\dag
}]=\delta _{KL}
\end{eqnarray}%
These include qubit optical, Zeeman (or hyperfine) coherences and population
operators, as well as bosonic annihilation, creation operators for the
cavity, spontaneous emission, cavity decay and the CM vibrational
modes.\medskip

The Hamiltonians involve certain coupling constants defined as follows:%
\begin{eqnarray}
\Omega _{ia} &=&-i\sum_{c}\sqrt{\frac{\omega _{c}}{2\epsilon _{0}\hbar V}}%
(d_{2a}\cdot u_{c})\alpha _{c}\exp (ik_{c}\cdot r_{i0}-\omega _{c}t) \\
g_{ia} &=&-i\sqrt{\frac{\omega _{b}}{2\epsilon _{0}\hbar V_{b}}}(d_{2a}\cdot
u_{b})\exp (ik_{b}\cdot r_{i0}) \\
g_{k}^{ia} &=&-i\sqrt{\frac{\omega _{k}}{2\epsilon _{0}\hbar V}}(d_{2a}\cdot
u_{k})\exp (ik\cdot r_{i0}) \\
p_{K}^{ia} &=&\sqrt{\frac{\omega _{b}}{2\epsilon _{0}\hbar V_{b}}}\sqrt{%
\frac{\hbar }{2m\nu _{K}}}(d_{2a}\cdot u_{b})(k_{b}\cdot S_{iK})\times 
\nonumber \\
&&\times \exp (ik_{b}\cdot r_{i0}) \\
\Theta _{K}^{ia} &=&\sum_{c}\sqrt{\frac{\omega _{c}}{2\epsilon _{0}\hbar V}}%
\sqrt{\frac{\hbar }{2m\nu _{K}}}(d_{2a}\cdot u_{c})(k_{c}\cdot S_{iK})\times
\nonumber \\
&&\times \alpha _{c}\exp i(k_{c}\cdot r_{i0}-\omega _{c}t) \\
t_{K}^{i} &=&ie_{T}\frac{1}{\sqrt{2\epsilon _{0}\omega _{b}V_{b}}}\sqrt{%
\frac{\nu _{K}}{2m}}(k_{b}\cdot S_{iK})\exp (ik_{b}\cdot r_{i0})
\end{eqnarray}%
where $a=0,1$\ unless stated otherwise. The cavity mode volume is $V_{b}$,
the SE mode volume $V$. Each ion has charge $e_{T}$.\medskip

\subsection{Dynamics and\ Decoherence Time\protect\medskip}

The total density operator for the qubits, ancilla and reservoirs satisfies
the Liouville-von Neumann equation 
\begin{eqnarray}
i\hbar \frac{\partial W}{\partial t} &=&[H,W] \\
H &=&H_{S}+V_{S}+H_{R}+V_{I}
\end{eqnarray}%
\medskip

The initial condition is given by%
\begin{equation}
W(0)=\rho _{S}(0)\rho _{R}(0)
\end{equation}%
and represents an uncorrelated state for qubits and reservoirs, the qubits
initially being in a pure state $|\psi _{S}\rangle $ and $\rho _{S}(0)=|\psi
_{S}\rangle \langle \psi _{S}|,$whilst the reservoirs are in thermal
states.\medskip

The reduced density operator for qubits and ancilla is defined as%
\begin{equation}
\rho _{S}=Tr_{R}W
\end{equation}%
and its general evolution allows for both coherent coupling and reservoir
interactions.\medskip

For coherent evolution due to $V_{S}$\ only, the reduced density operator
for qubits and ancilla would satisfy 
\begin{equation}
i\hbar \frac{\partial \rho _{S0}}{\partial t}=[H_{S}+V_{S},\rho _{S0}]
\end{equation}%
where the same initial condition $\rho _{S0}(0)=\rho _{S}(0)$\ can be chosen
as for the general evolution.\medskip\ 

The fidelity is defined by 
\begin{equation}
F=Tr_{S}\rho _{S0}\rho _{S}
\end{equation}%
and compares the actual and idealised evolution of the qubit system. \medskip

The decoherence timescale is defined as the time scale over which actual
quantum computer evolution (non-unitary) differs\ significantly from
idealised coherent evolution (unitary). The decoherence time scale\ will be
defined via the time dependence of the fidelity. It is related to certain
basic time scales due to qubit-environment coupling. These are: (a) the
reservoir correlation time $\tau _{c}$ - for EM field SE modes $\tau
_{c}\thicksim 10^{-17}$s (b) Markovian relaxation times - $T_{1}$ for
populations, $T_{2}$ for coherences - for EM field SE modes $%
T_{1,2}\thicksim 10^{-8}$s (c) the thermalisation time $\tau _{b}\,=\hbar
/k_{B}T$ - At $1\mu K$\ $\tau _{b}\thicksim 10^{-5}$s. The decoherence time
scale will also depend on factors such as qubit system state, the numbers of
qubits and the reservoirs involved.\medskip

Markovian evolution occurs for\ $t\gg \tau _{c}$\ and the reduced density
operator satisfies a master equation%
\begin{eqnarray}
i\hbar \frac{\partial \rho _{S}}{\partial t} &=&[H_{S}+V_{S},\rho
_{S}]+L\rho _{S} \\
L\rho _{S} &=&-i\tsum\limits_{ab}\Delta _{ab}[S_{a}S_{b}^{\dag },\rho _{S}] 
\nonumber \\
&&+\tsum\limits_{ab}\Gamma _{ab}\{[S_{b}^{\dag },\rho
_{S}S_{a}]+[S_{b}^{\dag }\rho _{S},S_{a}]\}.
\end{eqnarray}%
The master equation includes the Liouville superoperator term involving
Markovian relaxation $\Gamma _{ab}$ and frequency shift $\Delta _{ab}$
matrices and system operators $S_{a}$, where the system-reservoir
interaction $V_{I}$ is sum of products of system and reservoir operators%
\begin{eqnarray}
V_{I} &=&\tsum\limits_{a}S_{a}R_{a} \\
\lbrack H_{S},S_{a}] &=&\hbar \omega _{a}S_{a}.
\end{eqnarray}%
The Markovian evolution requires the reservoir correlation functions $%
\langle \widetilde{R_{a}}(t)\widetilde{R_{b}}(t-\tau )^{\dag }\rangle $
approach zero\ for $\tau \gg \tau _{c}$, and the Markovian relaxation $%
\Gamma _{ab}=\Gamma _{ba}^{\ast }$ and frequency shift $\Delta _{ab}=\Delta
_{ba}^{\ast }$ matrices are given via reservoir correlation functions as:

\begin{eqnarray}
C_{ab} &=&\tint\limits_{0}^{t}d\tau \langle \widetilde{R_{a}}(t)\widetilde{%
R_{b}}(t-\tau )^{\dag }\rangle \exp (-(i\omega _{b}+\epsilon )\tau ) \\
\Gamma _{ab} &=&(C_{ab}+C_{ba}^{\ast })/2,\quad \Delta
_{ab}=(C_{ab}-C_{ba}^{\ast })/2i
\end{eqnarray}%
\medskip\ \ \ \ \ \ 

In the Markovian intermediate time regime $(\tau _{d}\approx t\gg \tau _{c}) 
$ the rate $(\partial F/\partial t)_{0}\ $specifies the decoherence rate for
the qubit system, and its inverse defines the decoherence time $\tau _{D}$
The decoherence time can be expressed in terms of the Markovian relaxation
matrix elements and qubit system averages of products of fluctuation
operators\ 
\begin{equation}
\frac{1}{\tau _{D}}\equiv -\left( \frac{\partial F}{\partial t}\right)
_{t\gg \tau _{c}\rightarrow 0}=2\left( \tsum\limits_{ab}\Gamma _{ab}\langle
\Delta S_{a}\Delta S_{b}^{\dag }\rangle _{S}\right) _{t\gg \tau
_{c}\rightarrow 0}  \label{Eq.DecohTime}
\end{equation}%
Here the qubit and ancilla are in pure state $|\chi _{S}\rangle $,\ which
may differ from initial state $|\psi _{S}\rangle $ due to gating processes.
For $t\gg \tau _{c}\rightarrow 0$ we take $\rho _{S0}=\rho _{S}=|\chi
_{S}\rangle \langle \chi _{S}|$. In the above $\Delta S_{a}=S_{a}-\langle
S_{a}\rangle _{S}$ and $\langle \Omega \rangle _{S}\equiv Tr_{S}(\Omega \rho
_{S})$. The expression for the decoherence time scale is quite general and
may have applications for decoherence in other macroscopic systems, such as
Bose condensates or superconductors or in the theory of quantum measurement.
A special case of this result is given in Ref. \cite{Barnett96a}, which
deals with Bose condensates.\medskip

Expressions can be obtained for the loss of fidelity. During a time $T$ much
smaller than $\tau _{D}$, the change in fidelity if no gating is occuring is

\begin{equation}
\Delta F=-(\frac{1}{\tau _{D}})_{no-gating}.T
\end{equation}%
and if gating is taking place during a time $\Delta T$\ the change in
fidelity is

\begin{equation}
\Delta F=-(\frac{1}{\tau _{D}})_{gating}.\Delta T
\end{equation}%
For idealised quantum computation we require $\Delta F\ll 1$.\medskip

In the short-time regime $(t\ll \tau _{c})$\ the time dependent fidelity may
be written in a power series \cite{Dalton03a}%
\begin{eqnarray}
F(t) &=&1-(\frac{t}{\tau _{1}})-(\frac{t^{2}}{2\tau _{2}^{2}})+.. \\
\frac{\hbar }{\tau _{1}} &=&0 \\
\frac{\hbar ^{2}}{2\tau _{2}^{2}} &=&Tr_{BC}\langle \psi
_{S}|V_{I}(0)^{2}|\psi _{S}\rangle \rho _{B}(0)\rho _{C}(0)  \nonumber \\
&&-Tr_{BC}\langle \psi _{S}|V_{I}(0)|\psi _{S}\rangle ^{2}\rho _{B}(0)\rho
_{C}(0) \\
&\equiv &Tr_{BC}(\langle \Delta V_{I}(0)^{2}\rangle _{S})\rho _{B}(0)\rho
_{C}(0)
\end{eqnarray}%
\medskip The qubit and ancilla are initially in a pure state $|\psi
_{S}\rangle $ and $\rho _{S}(0)=|\psi _{S}\rangle \langle \psi _{S}|$. Here $%
\Delta V_{I}(0)=V_{I}(0)-\langle V_{I}(0)\rangle _{S}$ where $\langle \Omega
\rangle _{S}\equiv Tr_{S}(\Omega \rho _{S})$. The times $\tau _{1},\tau
_{2},..$specify characteristic decoherence times for the qubit and ancilla
system, their inverses defining decoherence rates. For the uncorrelated
initial state $W(0)$, only $\tau _{2}$ is involved in specifying the short
time decoherence. The decoherence time scale $\tau _{2}$\ depends on qubit
and reservoir averages of the fluctuation in the system-reservoir
interaction operator squared. The quadratic time dependence of the fidelity
is characteristic of the quantum Zeno effect. However, results obtained for $%
\tau _{2}$\ are due to non RWA terms in the system-reservoir coupling, and
may merely reflect the artificial nature of the uncorrelated
system-reservoir initial state.\bigskip

\section{Results}

\subsection{Case of\ No\ Gating\textbf{\protect\medskip }}

For the situation where no gating processes are occuring, there is no upper
state $|2\rangle $ amplitude, the cavity mode is in a no photon state $%
|0\rangle _{A}$\ and the qubit-ancilla state $|\chi _{S}\rangle $ is given
by $|\phi _{Q}\rangle |0\rangle _{A}$, where qubit system state is $|\phi
_{Q}\rangle $. This situation corresponds to states produced after idealised
coherent gating processes. To examine unfavourable scenarios, the reservoir
temperature $T$ is assumed non-zero and the qubits and the cavity is assumed
to have a low $Q$, so that spontaneous emission decay leads to a larger
decoherence rate than would otherwise be the case. Spontaneous emission is
the dominent decoherence process, and only its contribution is shown.\medskip

In the Markovian intermediate time regime $(\tau _{d}\approx t\gg \tau _{c})$
the decoherence time is given by%
\begin{equation}
\frac{1}{\tau _{D}}=\exp (-\frac{\hbar \omega _{0}}{k_{B}T})\tsum\limits_{ab}%
\sqrt{\Gamma _{a}\Gamma _{b}}\cos \theta _{ab}\tsum\limits_{i}\langle \sigma
_{ab}^{i}\rangle
\end{equation}%
The expression involves Zeeman (or hyperfine) coherences for $i$th qubit $%
\langle \sigma _{ab}^{i}\rangle \;(a\neq b)\;(a,b=0,1)$\ and populations for 
$i$th qubit $\langle \sigma _{aa}^{i}\rangle \;(a=0,1)$. The optical
transition frequencies are $\omega _{2a}\thicksim \omega _{0}\;(a=0,1)$, the
spontaneous emission decay rates for 2-a transition are $\Gamma _{a}$\ and
the angle between dipole matrix elements $d_{2a},d_{2b}$ is $\theta _{ab}$%
.\medskip

For the case of the Hadamard state (uncorrelated), the qubit state is $|\phi
_{Q}\rangle =\tprod\limits_{i}|\phi _{Q}\rangle _{{\small i}}$, where $i$ th
qubit state vector is $|\phi _{Q}\rangle _{{\small i}}=(|0\rangle _{{\small i%
}}+|1\rangle _{{\small i}})/2^{{\small 1/2}}$, we find that

\begin{equation}
\frac{1}{\tau _{D}}=\frac{1}{2}N\exp (-\frac{\hbar \omega _{0}}{k_{B}T}%
)\tsum\limits_{ab}\sqrt{\Gamma _{a}\Gamma _{b}}\cos \theta _{ab}
\end{equation}%
and note that theecoherence time for Hadamard state can be infinite for the
lambda qubit system if choose $d_{20}+d_{21}=0.$\medskip

On the other hand, for the case of the GHZ\ state (correlated), where the
qubit system state vector is $|\phi _{Q}\rangle =(|00..0\rangle
+|11..1\rangle )/2^{{\small 1/2}}$, we obtain

\begin{equation}
\frac{1}{\tau _{D}}=\frac{1}{2}N\exp (-\frac{\hbar \omega _{0}}{k_{B}T}%
)\tsum\limits_{a}\Gamma _{a}
\end{equation}%
Here the decoherence time scale is still very long, due to the upper state
Boltzmann factor. With $N\thickapprox $10$^{4}$\ qubits and $\Gamma
\thickapprox $10$^{8}$s$^{-1}$, $\omega _{0}\thickapprox $10$^{15}$s$^{-1}$, 
$T\ \thickapprox $300 $K$, we find $\tau _{D}\thickapprox $10$^{19}$%
s.\medskip

Overall, spontaneous emission due to electric dipole coupling is dominent
cause of decoherence. Other terms such as Lamb-Dicke coupling can be
ignored, so the qubits can be treated as stationary. The decoherence time
scales as $1/N$ for this intermediate time regime but it is very long. The
present results for this case of memory decoherence are consistent with
those of Garg \cite{Garg96a}, who also found very low decoherence rates
allowing for vibrational motion in Cirac-Zoller type quantum computers \cite%
{Cirac95a} with large numbers of qubits.\medskip 

\subsection{Case of\ One Qubit Gating\textbf{\protect\medskip }}

For the one qubit gating process with two-photon resonant Raman gating
fields detuned from one photon resonance, the upper state $|2\rangle $
amplitude for $i$th (gated) qubit becomes slightly non-zero, though the
cavity mode remains in the zero photon state $|0\rangle _{A}$ $\ |\chi
_{S}\rangle $ given by $|\phi _{Q}\rangle |0\rangle _{A}$, with the qubit
state $|\phi _{Q}\rangle $\ \ $(a_{j}=0,1)$\ given as

\begin{equation}
{\small |\phi }_{Q}{\small \rangle =}\sum_{\{a\}}C_{0}({\small \{a\}})%
{\small |\{a\}\rangle +}\sum_{\{a_{i}\}}C_{0}({\small \{a}_{i}{\small \};2}%
_{i}){\small |\{a}_{i}{\small \};2}_{i}{\small \rangle }
\end{equation}%
Here $\{a\}\equiv \{a_{1},a_{2},..,a_{N}\}$, $\{a_{i}\}\equiv
\{a_{1},.,a_{i-1},a_{i+1},..\}.$However spontaneous emission decay is small
in the high Q cavity. The reservoir temperature is assumed zero.\medskip

In the Markovian intermediate time regime $(\tau _{d}\approx t\gg \tau _{c})$
the decoherence time is given by

\begin{eqnarray}
\frac{1}{\tau _{D}} &=&2\{\tsum\limits_{k\neq i}\tsum\limits_{ab}\langle
\sigma _{ab}^{k}\rangle \Gamma _{ka-C;\,kb-C}  \nonumber \\
&&+\tsum\limits_{ab}(\langle \sigma _{ab}^{i}\rangle -\langle \sigma
_{-}^{ia}\rangle \langle \sigma _{+}^{ib}\rangle )\Gamma _{ia-;\,ib-} 
\nonumber \\
&&+\tsum\limits_{ab}(\langle \sigma _{22}^{i}\rangle \delta _{ab}-\langle
\sigma _{+}^{ia}\rangle \langle \sigma _{-}^{ib}\rangle )\Gamma _{ia+;\,ib+}
\nonumber \\
&&+\tsum\limits_{ab}\langle \sigma _{ab}^{i}\rangle \Gamma
_{ia-C;\,ib-C}+\tsum\limits_{a}\langle \sigma _{22}^{i}\rangle \Gamma
_{ia+C;\,ia+C}\}
\end{eqnarray}%
The decoherence time involves Markovian relaxation elements and state
dependent quantities for the qubit system. For the gated qubit terms,
optical coherences $\langle \sigma _{\pm }^{ia}\rangle $, Zeeman (or
hyperfine) coherences $\langle \sigma _{ab}^{i}\rangle $, upper state
population $\langle \sigma _{22}^{i}\rangle $\ and lower state populations $%
\langle \sigma _{aa}^{i}\rangle $\ are involed. For the non-gated qubits $%
(k\neq i)$, Zeeman (or hyperfine) coherences $\langle \sigma
_{ab}^{k}\rangle $\ and lower state populations $\langle \sigma
_{aa}^{k}\rangle $\ are present.\medskip

Expressions for the zero temperature relaxation matrix elements\ are as
follows:

\begin{eqnarray}
\quad \Gamma _{ka-C;\,kb-C} &=&\frac{i}{8}\eta ^{2}g_{ka}g_{kb}^{\ast }\frac{%
\omega _{ab}}{\omega _{0}^{2}}\quad {\small (k\neq i,\,k=i)} \\
\Gamma _{ia-;\,ib-} &=&\frac{i}{2}\eta ^{2}(\Omega _{ia}^{\ast }\Omega
_{ib}^{\ast }-\Omega _{ia}\Omega _{ib})/\Delta _{ci}^{0} \\
\Gamma _{ia+;\,ib+} &=&\frac{1}{2}\sqrt{\Gamma _{a}\Gamma _{b}}\cos \theta
_{ab}  \nonumber \\
&&+\frac{i}{2}\eta ^{2}(\Omega _{ia}^{\ast }\Omega _{ib}^{\ast }-\Omega
_{ia}\Omega _{ib})/\Delta _{ci}^{0} \\
\Gamma _{ia+C;\,ib+C} &=&i\eta ^{2}g_{ia}g_{ib}^{\ast }\frac{{\small %
(1-\delta }_{ab}{\small )(-1)}^{a}}{\nu _{\max }}
\end{eqnarray}%
In these formulae the Lamb-Dicke (LD) parameter is $\eta $, the gating EM
field Rabi frequencies are $\Omega _{ia}$ $(a=0,1)$, the one photon
detunings for the Raman gating field are $\Delta _{ci}^{0}$\ and the one
photon Rabi frequencies for the cavity mode are $g_{ka}$ $(a=0,1)$. The
Zeeman (or hyperfine) transition frequencies are $\omega _{ab}\;(a\neq
b)\;(a,b=0,1)$. Vibrational frequencies range from zero up to $\nu _{\max }$%
, and the vibrational modes have zero phonons at absolute zero. In the case
studied the cavity mode is resonant with the 2-1 transition $\omega
_{b}=\omega _{21}\sim \omega _{0}$, but both one photon Rabi frequencies $%
g_{ka}$\ are assumed non-zero. Approximations based on $\Delta
_{ci}^{0}\;\gg \nu _{\max }$, $\Gamma $\ and $\omega _{10}\;\gg \nu _{\max }$%
\ have been used.\medskip

The populations and coherences are found for large one photon detuning,
assuming $\Omega _{i0}=\Omega _{i1}\exp i\Delta \phi =\Omega (t)$, where the
common amplitude $\Omega $\ has a maximum $\Omega _{m}$\ and a width $\Delta
T$. For a gated qubit initially in state $|0\rangle $\ (see Vitanov et al 
\cite{Vitanov97a}) we find that

\begin{eqnarray}
\langle \sigma _{01}^{i}\rangle &=&\langle \sigma _{10}^{i}\rangle ^{\ast
}=i\sin \theta .\cos \theta .\exp i\Delta \phi \\
\langle \sigma _{00}^{i}\rangle &=&\cos ^{2}\theta ,\quad \langle \sigma
_{11}^{i}\rangle =\sin ^{2}\theta \\
\langle \sigma _{+}^{i0}\rangle &=&\langle \sigma _{-}^{i0}\rangle ^{\ast }=-%
\frac{\Omega }{\Delta }.\exp (-i\theta ).\cos \theta \\
\langle \sigma _{+}^{i1}\rangle &=&\langle \sigma _{-}^{i1}\rangle ^{\ast }=-%
\frac{\Omega }{\Delta }.\exp (-i\theta ).i\sin \theta .\exp i\Delta \phi \\
\langle \sigma _{22}^{i}\rangle &=&\Omega ^{2}/\Delta ^{2}
\end{eqnarray}%
In these expressions the quantity $\theta (t)\ $is defined by the integral
of the two photon Rabi frequency $\Omega _{R}=\Omega ^{2}/\Delta $\ 
\begin{equation}
\theta (t)=\tint\limits_{-\infty }^{t}dt^{\prime }\Omega (t^{\prime
})^{2}/\Delta
\end{equation}%
The gating time $\Delta T$\ is given by $\theta =\pi /2$ for $t\simeq \Delta
T$, corresponding to the time the qubit takes to evolve into state $%
|1\rangle $\ \ 
\begin{equation}
\Delta T\simeq \frac{\pi }{2}\frac{\Delta }{\Omega _{m}^{2}}
\label{Eq.OneQBitGateTime}
\end{equation}%
\medskip

The parameters used in both the one and two qubit gating processes are set
out in Table 1\medskip

\begin{center}
\textbf{Table 1}. Parameters used in gating processes
\end{center}

\begin{tabular}{|l||l|}
\hline
$\omega _{b}\approx \omega _{0}$ & $3.10^{15}$s$^{-1}$ \\ \hline
$\omega _{10}$ & $6.10^{9}$s$^{-1}$ \\ \hline
$\Delta _{ci}^{0}$ & $3.10^{10}$s$^{-1}$ \\ \hline
$\nu _{\max }$ & $8.10^{7}$s$^{-1}$ \\ \hline
\end{tabular}%
\quad 
\begin{tabular}{|l||l|}
\hline
$\left\vert \Omega _{ia}\right\vert $ & $3.10^{6}$s$^{-1}$ \\ \hline
& $3.10^{8}$s$^{-1}$ \\ \hline
$g_{ka}$ & $3.10^{8}$s$^{-1}$ \\ \hline
$\Gamma _{a}$ & $3.10^{4}$s$^{-1}$ \\ \hline
$\eta $ & $6.10^{-2}$ \\ \hline
\end{tabular}%
\quad 
\begin{tabular}{|l||l|}
\hline
$\Gamma _{cav}$ & $3.10^{8}$s$^{-1}$ \\ \hline
$Q$ & $1.10^{7}$ \\ \hline
\end{tabular}

These correspond to optical and hyperfine transitions with a high Q optical
cavity in the medium coupling regime. As discussed previously, these
parameters are chosen to be the same as in the work of Tregenna et al \cite%
{Tregenna02a}, so that their model\ can be compared to the present one where
decoherence effects due to vibrational motion is allowed for as qubit
numbers increase. The cavity coupling constant and cavity decay rate are
made equal, and large compared to spontaneous emission (SE) rate, and the
gating field one photon Rabi frequency is large compared to the SE decay
rate \ Two cases are studied, corresponding to the gating field one photon
Rabi frequency being (i) small compared to (ii) the same as the cavity
coupling constant and cavity decay rate. Case (i) and with zero $g_{k0}$\
applies in Tregenna et al \cite{Tregenna02a}. The maximum vibration
frequency and Lamb-Dicke parameter are calculated for a 3D Ca$^{+}$lattice
with lattice spacing $3\mu $. A standard approach to the theory of lattice
vibrations is used \cite{Ziman65a}, in which the vibrational potential
energy is given by a quadratic form of the small displacements of the qubits
from the lattice sites, as in Eq.\ref{Eq.HamLattVib}. The interaction
between each pair of ionic qubits is obtained from electrostatics, from
which the $V_{ij}^{\alpha \beta }$\ are obtained. For the cubic lattice
case, the vibration frequencies are obtained from Eq.\ref{Eq.LatticeVibFreq}%
, and are approximately proportional to the magnitude of the wave vector for
each vibrational mode. Expressions for the quantities $S_{i\alpha ;K}$\ are
obtained from Eq.\ref{Eq.LatticeVibFreq}.$\medskip $

Using these parameters, the zero temperature relaxation elements\ can be
calculated and the results are given in Table 2.\medskip

\begin{center}
\textbf{Table 2}. Relaxation elements in one qubit gating process
\end{center}

\begin{tabular}{|l|l|l|l|}
\hline
$\Gamma _{ka-C;\,kb-C}$ & $\Gamma _{ia-;\,ib-}$ & $\Gamma _{ia+;\,ib+}$ & $%
\Gamma _{ia+C;\,ib+C}$ \\ \hline\hline
$3.10^{{\small -9}}$s$^{{\small -1}}$ & $1.10^{{\small 0}}$s$^{{\small -1}%
},1.10^{{\small 4}}$s$^{{\small -1}}$ & $3.10^{{\small 4}}$s$^{{\small -1}}$
& $4.10^{{\small 6}}$s$^{{\small -1}}$ \\ \hline
&  & $1.10^{{\small 0}}$s$^{{\small -1}},1.10^{{\small 4}}$s$^{{\small -1}}$
&  \\ \hline
\end{tabular}%
\smallskip \medskip

The overall contributions to the decoherence rate can be obtaining by
combining the qubit factor (for $\theta \simeq \pi /4$, midway through
process) and the relaxation matrix elements for the various terms $(1-5)$\
involved in the expression for the decoherence rate $1/\tau _{D}$. The
results are presented in Table 3. Here NG, G refer to non-gated and gated
qubits respectively. The two gating field cases are: (i) $\Omega =3.10^{6}$s$%
^{{\small -1}}$\ (ii) $\Omega =3.10^{8}$s$^{{\small -1}}.$The corresponding
gating times are $10^{-2}$s, $10^{-6}$s (see Eq.\ref{Eq.OneQBitGateTime}),
which are long compared to the correlation time for the vibrational modes
reservoir of $\ 10^{-7}$s (given by $2\pi /\nu _{\max }$), and much longer
than the correlations times associated with the SE or cavity decay modes
reservoirs.\medskip

\begin{center}
\textbf{Table 3}. Contributions to decoherence rate in one qubit gating
process
\end{center}

\begin{tabular}{|l|l|l|l|l|}
\hline
Term & Relaxation & Relaxation & Qubit & Overall $1/\tau _{D}$ \\ \hline
& Effect Due & Rate & Factor & Contribution \\ \hline\hline
$1,NG$ & $LD-Cavity$ & $3.10^{-9}$s$^{-1}$ & $0.5$ & $N.2.10^{-9}$s$^{-1}$
\\ \hline
$2,G$ & $LD-Gating$ & $1.10^{0}$s$^{-1}$ & $0.5$ & $(i)\;5.10^{-1}$s$^{-1}$
\\ \hline
&  & $1.10^{4}$s$^{-1}$ & $0.5$ & $(ii)\;5.10^{+3}$s$^{-1}$ \\ \hline
$3,G$ & $SE$ & $3.10^{4}$s$^{-1}$ & $10^{-8}$ & $(i)\;3.10^{-4}$s$^{-1}$ \\ 
\hline
&  & $3.10^{4}$s$^{-1}$ & $10^{-4}$ & $(ii)\;3.10^{0}$s$^{-1}$ \\ \hline
$4,G$ & $LD-Cavity$ & $3.10^{-9}$s$^{-1}$ & $0.5$ & $1.10^{-9}$s$^{-1}$ \\ 
\hline
$5,G$ & $LD-Cavity$ & $4.10^{6}$s$^{-1}$ & $10^{-8}$ & $(i)\;4.10^{-2}$s$%
^{-1}$ \\ \hline
&  & $4.10^{6}$s$^{-1}$ & $10^{-4}$ & $(ii)\;4.10^{+2}$s$^{-1}$ \\ \hline
\end{tabular}%
\medskip

The scaling of the decoherence rate with qubit number for tne case of one
qubit Raman gating gives an overall decoherence time which is essentially
independent of $N$. This is because the terms for gated qubit $i$\ do not
scale with $N$, whilst the non-gated qubit contributions (which scale with $%
N $) are negligible in comparison, even for $N\sim 10^{4}$\ qubits. These
features are clear from the results in Table 3. For both the small and
larger gating field cases, the largest overall contribution to the
decoherence rate arises from the terms associated with Lamb-Dicke coupling
of the gated qubit with the gating field $\tsum\limits_{ab}(\langle \sigma
_{ab}^{i}\rangle \,-\langle \sigma _{-}^{ia}\rangle \langle \sigma
_{+}^{ib}\rangle )\Gamma _{ia-;\,ib-}$. The next largest overall
contribution arises from the terms associated with Lamb-Dicke coupling of
the gated qubit with the cavity mode $\tsum\limits_{a}\langle \sigma
_{22}^{i}\rangle \Gamma _{ia+C;\,ia+C}$. The larger relaxation rate $\Gamma
_{ia+C;\,ia+C}$\ for the latter is balanced by the smaller qubit factor $%
\langle \sigma _{22}^{i}\rangle $\ associated with the upper state
population.\medskip

The overall fidelity loss can then be obtained. Since terms associated with
Lamb-Dicke coupling with the gating field are dominent, these terms combined
with the gating time gives for the overall fidelity loss%
\begin{eqnarray}
\Delta F &\simeq &-\frac{1}{2}.\frac{\eta ^{2}\Omega _{m}^{2}}{\Delta }.%
\frac{\pi }{2}\frac{\Delta }{\Omega _{m}^{2}} \\
&\simeq &-\frac{\pi }{4}\eta ^{2}
\end{eqnarray}%
We note that the fidelity loss only depends on Lamb-Dicke parameter. For $%
\eta =6.10^{-2}$, the fidelity loss is $2.10^{-3}$. This is reasonably small
though still somewhat large for fault-tolerant quantum computation - see 
\cite{Preskill99a}, \cite{LANL04a}.\medskip

\subsection{Case of\ Two Qubit Gating\protect\medskip}

For the two qubit gating process, with resonant gating fields coupled to the
2-1 transition for $i$th (control) qubit, and coupled to the 2-0 transition
for the $j$th (target) qubit, and the cavity mode resonant with 2-1
transition, but uncoupled to the 2-0 transition \cite{Tregenna02a}, the
upper state $|2\rangle $ amplitude for the $i$th (control) and $j$th
(target) gated qubits could be non-zero. The cavity mode is in zero photon
state $|0\rangle _{A}$ when one or both gated qubits $i,j$ are in the upper
state, but could be in one photon state $|1\rangle _{A}\ $when both are in
lower states. Here $|\chi _{S}\rangle $ given by $|\phi _{Q}^{0}\rangle
|0\rangle _{A}+|\phi _{Q}^{1}\rangle |1\rangle _{A}$, with qubit states $%
|\phi _{Q}^{0,1}\rangle $

\begin{eqnarray}
{\small |\phi }_{Q}^{0}{\small \rangle } &{\small =}&\sum_{\{a_{ij}\}}C_{0}(%
{\small \{a}_{ij}{\small \};2}_{i}{\small 2}_{j}){\small |\{a}_{ij}{\small %
\};2}_{i}{\small 2}_{j}{\small \rangle +}\sum_{\{a_{i}\}}C_{0}({\small \{a}%
_{i}{\small \};2}_{i}){\small |\{a}_{i}{\small \};2}_{i}{\small \rangle } 
\nonumber \\
&&{\small +}\sum_{\{a_{j}\}}C_{0}({\small \{a}_{j}{\small \};2}_{j}){\small %
|\{a}_{j}{\small \};2}_{j}{\small \rangle +}\sum_{\{a\}}C_{0}({\small \{a\}})%
{\small |\{a\}\rangle } \\
{\small |\phi }_{Q}^{1}{\small \rangle } &{\small =}&\sum_{\{a\}}C_{1}(%
{\small \{a\}}){\small |\{a\}\rangle .}
\end{eqnarray}%
Here $\{a\}\equiv \{a_{1},a_{2},..,a_{N}\}$, $\{a_{i}\}\equiv
\{a_{1},.,a_{i-1},a_{i+1},..\}$, $\{a_{j}\}\equiv
\{a_{1},.,a_{j-1},a_{j+1},..\}$, $\{a_{ij}\}\equiv
\{a_{1},.,a_{i-1},a_{i+1},..,a_{j-1},a_{j+1},..\}$, and $(a_{i},a_{j}=0,1)$.
However spontaneous emission decay is small in the high Q cavity. The
reservoir temperature is assumed zero.\medskip

In the Markovian intermediate time regime $(\tau _{d}\approx t\gg \tau _{c})$
the decoherence time is given as the sum of non-gated and gated qubit
contributions as%
\begin{equation}
\left( \frac{1}{\tau _{D}}\right) =\left( \frac{1}{\tau _{D}}\right)
_{NG}+\left( \frac{1}{\tau _{D}}\right) _{G}.
\end{equation}%
The non-gated (NG) qubits contribution involves 7 terms 
\begin{eqnarray}
\left( \frac{1}{\tau _{D}}\right) _{NG} &=&2\{\tsum\limits_{k\neq
i,j}\tsum\limits_{ab}\langle \sigma _{ab}^{k}\rangle \Gamma _{ka-;\,kb-} 
\nonumber \\
&&+\tsum\limits_{k\neq i,j}\tsum\limits_{ab}(\langle \sigma _{ab}^{k}b^{\dag
}\rangle \Gamma _{ka-;\,kb-C}+cc)  \nonumber \\
&&+\tsum\limits_{k\neq i,j}\tsum\limits_{ab}(\langle \sigma
_{ab}^{k}b\rangle \Gamma _{ka-;\,kb-C+}+cc)  \nonumber \\
&&+\tsum\limits_{k\neq i,j}\tsum\limits_{ab}\langle \sigma _{ab}^{k}bb^{\dag
}\rangle \Gamma _{ka-C;\,kb-C}  \nonumber \\
&&+\tsum\limits_{k\neq i,j}\tsum\limits_{ab}\langle \sigma _{ab}^{k}b^{\dag
}b\rangle \Gamma _{ka-C+;\,kb-C+}\}.
\end{eqnarray}%
The gated qubits contribution consist of 120 terms. Some of the terms are

\begin{eqnarray}
\left( \frac{1}{\tau _{D}}\right) _{G}^{(23)} &=&2\tsum\limits_{ab}(\langle
\sigma _{ab}^{i}\rangle -\langle \sigma _{-}^{ia}\rangle \langle \sigma
_{+}^{ib}\rangle )\Gamma _{ia-;\,ib-} \\
\left( \frac{1}{\tau _{D}}\right) _{G}^{(24)} &=&2\tsum\limits_{ab}(\langle
\sigma _{a2;\,2b}^{ij}\rangle -\langle \sigma _{-}^{ia}\rangle \langle
\sigma _{+}^{jb}\rangle )\Gamma _{ia-;\,jb-} \\
\left( \frac{1}{\tau _{D}}\right) _{G}^{(34)} &=&2\tsum\limits_{ab}(\langle
\sigma _{a2;\,2b}^{ji}\rangle -\langle \sigma _{-}^{ja}\rangle \langle
\sigma _{+}^{ib}\rangle )\Gamma _{ja-;\,ib-} \\
\left( \frac{1}{\tau _{D}}\right) _{G}^{(35)} &=&2\tsum\limits_{ab}(\langle
\sigma _{ab}^{j}\rangle -\langle \sigma _{-}^{ja}\rangle \langle \sigma
_{+}^{jb}\rangle )\Gamma _{ja-;\,jb-} \\
\left( \frac{1}{\tau _{D}}\right) _{G}^{(120)} &=&2(\langle b^{\dag
}b\rangle -\langle b^{\dag }\rangle \langle b\rangle )\Gamma _{C+;\,C+}
\end{eqnarray}%
Some of these are equivalent to those for one qubit gating, others involve
different expressions, since the gating fields are now resonant rather than
having a large detuning. In addition, there are many new terms only present
for two qubit gating. Again, the decoherence time involves Markovian
relaxation elements and state dependent quantities for qubit system. The
terms for non-gated qubits $(k\neq i,j)$ - involve Zeeman (or hyperfine)
coherences $\langle \sigma _{ab}^{k}\rangle $\ and lower state populations $%
\langle \sigma _{aa}^{k}\rangle $, as well as quantities also involving the
cavity mode operators $\langle \sigma _{ab}^{k}b\rangle ,\langle \sigma
_{ab}^{k}b^{\dag }\rangle ,\langle \sigma _{ab}^{k}bb^{\dag }\rangle
,\langle \sigma _{ab}^{k}b^{\dag }b\rangle ,\langle \sigma _{aa}^{k}b\rangle
,\langle \sigma _{aa}^{k}b^{\dag }\rangle ,\langle \sigma _{aa}^{k}bb^{\dag
}\rangle $\ and $\langle \sigma _{aa}^{k}b^{\dag }b\rangle .$\ For the terms
involving gated qubits, optical coherences $\langle \sigma _{\pm
}^{ga}\rangle $,\ Zeeman (or hyperfine) coherences $\langle \sigma
_{ab}^{g}\rangle $,\ upper state populations $\langle \sigma
_{22}^{g}\rangle $,\ lower state populations $\langle \sigma
_{aa}^{g}\rangle $\ ($g$\ refers to $i$, $j$), as well as two qubit state
transitions $\langle \sigma _{22;\,ab}^{gh}\rangle $, $\langle \sigma
_{ab;\,22}^{gh}\rangle $, $\langle \sigma _{a2;\,2b}^{gh}\rangle $\ and $%
\langle \sigma _{2a;\,b2}^{gh}\rangle $\ are all involved.\ The two qubit
transitions are $\sigma _{ab;\,cd}^{gh}=(|a\rangle \langle c|)_{g}(|b\rangle
\langle d|)_{h}$, where $g\neq h$\ refers to $i\neq j$). In addition, there
are quantities involving cavity mode operators also, similar to those for
the non-gated qubits.\medskip

For the non-gated qubits, all relaxation elements $\Gamma _{ka-;\,kb-}$, $%
\Gamma _{ka-;\,kb-C}$, $\Gamma _{ka-;\,kb-C+}$, $\Gamma _{ka-C;\,kb-C}$\ and 
$\Gamma _{ka+C;\,kb+C}$\ are zero for the specific gating process \cite%
{Tregenna02a} involved. In particular, the last two are zero because the
cavity mode is resonantly coupled to the 2-1 transition and uncoupled to the
2-0 transition, as may be seen from the following expressions:

\begin{eqnarray}
\Gamma _{ka-C;\,kb-C} &=&\frac{i}{8}\eta ^{2}g_{ka}g_{kb}^{\ast }\frac{%
\omega _{ab}}{\omega _{0}^{2}} \\
\Gamma _{ka+C;\,kb+C} &=&i\eta ^{2}g_{ka}g_{kb}^{\ast }\frac{{\small %
(1-\delta }_{ab}{\small )(-1)}^{a}}{\nu _{\max }}
\end{eqnarray}%
\medskip

For the gated qubits ($i$\ control, $j$\ target qubit) some of the
relaxation elements are

\begin{eqnarray}
\Gamma _{ia-;\,ib-} &=&i\,\delta _{a1}\delta _{b1}\,\eta ^{2}(\Omega
_{i1}\Omega _{i1}-\Omega _{i1}^{\ast }\Omega _{i1}^{\ast })/\nu _{\max } \\
\Gamma _{ia-;\,jb-} &=&i\,\delta _{a1}\delta _{b0}\,\eta ^{2}\,(\widehat{k}%
_{ci}\cdot \widehat{k}_{cj})(\Omega _{i1}\Omega _{j0}-\Omega _{i1}^{\ast
}\Omega _{j0}^{\ast })/\nu _{\max }  \nonumber \\
&&\times \frac{1}{x_{ij}}\tint\limits_{0}^{x_{ij}}dx\,\frac{\sin x}{x} \\
\Gamma _{ja-;\,ib-} &=&(\Gamma _{ia-;\,jb-})^{\ast } \\
\Gamma _{ja-;\,jb-} &=&i\,\delta _{a0}\delta _{b0}\,\eta ^{2}(\Omega
_{j0}\Omega _{j0}-\Omega _{j0}^{\ast }\Omega _{j0}^{\ast })/\nu _{\max } \\
\Gamma _{C+;\,C+} &=&\frac{1}{2}\Gamma _{cav}.
\end{eqnarray}%
The quantity $x_{ij}=\sqrt{{\small 3}}a\,/\,\left\vert
r_{i0}-r_{j0}\right\vert $\ relates the qubit separation to the lattice size
\ As stated above, the cavity mode is resonant with 2-1 transition $(\omega
_{b}=\omega _{21}\sim \omega _{0})$, the cavity mode is uncoupled to the 2-0
transition $(g_{k0}=0)$, the control gating field is coupled to the 2-1
transition, the target gating field is coupled to the 2-0 transition and
both gating fields are on resonance. Approximations based on $\omega
_{0}\,\gg \omega _{10}\;\gg \,\nu _{\max }$\ are used in the derivations.
The vibrational modes have zero phonons at absolute zero.\medskip

\ Expressions for qubit populations, coherences and two qubit transitions
could be obtained based on the work of Tregenna et al \cite{Tregenna02a}.
Gating is based on the use of decoherence-free subspaces. States for the
gated qubits ($|0_{i}0_{j}\rangle $, $|0_{i}1_{j}\rangle $, $%
|1_{i}0_{j}\rangle $, $|1_{i}1_{j}\rangle $, $|A\rangle =(|1_{i}2_{j}\rangle
-|2_{i}1_{j}\rangle )/\sqrt{{\small 2}}$), with the cavity mode in zero
photon state $|0\rangle _{A}$\ are not directly coupled to one photon cavity
states (see \cite{Tregenna02a}) via the qubit-cavity interaction. A CNOT
gate can thus be performed with negligible cavity mode excitation, thereby
avoiding decoherence due to cavity mode decay. Their treatment assumes $%
\left\vert \Omega _{i1}\right\vert =$\ $\left\vert \Omega _{j0}\right\vert
=\Omega (t)$, where $\Omega $\ has a maximum $\Omega _{m}$\ and a width $%
\Delta T$. \medskip

The gating time is given by 
\begin{equation}
\Delta T\simeq \frac{2\pi }{\Omega _{m}}  \label{Eq.TwoQBiteGateTime}
\end{equation}

The parameters used for the two qubit gating case are the same as for one
qubit gating, except in accordance with Tregenna et al, the gating field is
weak, and only $\Omega _{m}=3.10^{6}$s$^{{\small -1}}$\ is used (see Table
1). The corresponding gating time is $2.10^{-6}$s (see Eq.\ref%
{Eq.TwoQBiteGateTime}), which is long compared to the correlation time for
the vibrational modes reservoir of $\ 10^{-7}$s (given by $2\pi /\nu _{\max
} $), and much longer than the correlations times associated with the SE or
cavity decay modes reservoirs.\smallskip Some relaxation elements obtained
are given in Table 4.\medskip

\begin{center}
\textbf{Table 4}. Relaxation elements in two qubit gating process
\end{center}

\begin{tabular}{|l|l|l|l|l|}
\hline
$\Gamma _{ia-;\,ib-}$ & $\Gamma _{ia-;\,jb-}=(\Gamma _{ja-;\,ib-})^{{\small %
\ast }}$ & $\Gamma _{ja-;\,jb-}$ & $\Gamma _{C+;\,C+}$ & $\Gamma
_{ka-C+;\,kb-C+}$ \\ \hline\hline
$4.10^{{\small 2}}$s$^{{\small -1}}$ & $4.10^{{\small 2}}$s$^{{\small -1}}$
& $4.10^{{\small 2}}$s$^{{\small -1}}$ & $1.10^{{\small 8}}$s$^{{\small -1}}$
& $0$\ s$^{{\small -1}}$ \\ \hline
&  &  &  & $1.10^{{\small 6}}$s$^{{\small -1}}$ \\ \hline
\end{tabular}%
\medskip

The scaling of the decoherence rate with qubit number for tne case of the
two qubit gating process treated in Tregenna et al \cite{Tregenna02a} gives
an overall decoherence time which is independent of $N$. This is because the
terms for gated qubits $i,j$\ do not scale with $N$, whilst the non-gated
qubit contributions (which scale with $N$) are zero for the present case
where the 2-1 transition resonantly coupled to cavity mode and the 2-0
transition is uncoupled. However, other parameter choices, such as having
both optical transitions coupled to the cavity mode, could lead to large
non-gated contributions, due to $\Gamma _{{\small ka-C+};\,{\small kb-C+}%
}\sim 1.10^{{\small +6}}$s$^{{\small -1}}$\ associated with LD coupling of
the qubits with the cavity and vibrational modes. If one photon is present,
so that $\langle \sigma _{ab}^{k}b^{\dag }b\rangle \sim 1$, then even modest
size qubit numbers $N\sim 10^{{\small 3}}$\ would lead to non-gated qubit
contributions exceeding those from other contributions, such as $\langle
b^{\dag }b\rangle \Gamma _{C+;\,C+}$, where $\Gamma _{C+;C+}\sim 1.10^{%
{\small +8}}$s$^{{\small -1}}$.\medskip

As for one qubit gating, the contributions from the various terms to $%
(1/\tau _{d})_{gating}$\ (and hence to the change in fidelity during the two
qubit gating period) are significantly different in size. The relaxation
element factor may be more important than the qubit state factor, and vice
versa. We would need to investigate all 120 terms to determine which
contribution is dominent. For example, for the terms numbered 23, 24, 34 and
35, the relaxation element is $\sim 4.10^{{\small 2}}$s$^{{\small -1}}$\ and
the qubit factors are $\sim 1$, giving a product $\sim 4.10^{{\small 2}}$s$^{%
{\small -1}}$. For term number 120, the relaxation element is $\sim 1.10^{%
{\small 8}}$s$^{{\small -1}}$\ and the qubit factor gives the probability $%
\langle b^{\dag }b\rangle $\ of finding one photon in the cavity mode. If
this probability is greater than $\sim 4.10^{{\small -6}}$\ (and it could be
as high as unity if decoherence free subspaces were not utilised during the
gating process), the term 120 would be more important than the terms 23, 24,
34 and 35. Term 120 is due to cavity decay. If the cavity decay term was the
most important, the reduction in fidelity would be given by

\begin{equation}
\Delta F=-(\langle b^{\dag }b\rangle -\langle b^{\dag }\rangle \langle
b\rangle )\,\Gamma _{cavity}\,\frac{2\pi }{\Omega _{m}}
\end{equation}%
which is $\sim 6.10^{{\small 2}}\,\langle b^{\dag }b\rangle $. The
probability of finding a photon in the cavity mode must be less than $10^{%
{\small -5}}$\ if the fidelity loss is to be reasonably small.\bigskip

\section{Discussion\protect\medskip}

The scaling of decoherence effects in circuit model quantum computers have
been studied for the situation where the number of qubits $N$\ becomes
large. Decoherence effects were specified via the fidelity, with its initial
rate of change defining the decoherence time scale. Expressions for the
decoherence time scale were obtained for the intermediate time regime via
Markovian theory. The general case was treated where the qubit system was in
any pure state, the reservoirs being in thermal states. The decoherence time
scale was expressed in terms of Markovian relaxation elements and
expectation values of products of fluctuation operators for the decohering
quantum system. The expression given in Eq.\ref{Eq.DecohTime} for the
decoherence time scale is quite general and may have applications for
treating decoherence in other macroscopic systems, such as Bose condensates
or superconductors or in quantum measurement theory.\medskip

A standard model involving $N$\ ionic qubits, each a three-state lambda
system, was studied, with localised, well-separated qubits undergoing
vibrational motion in a lattice of trapping potentials. Coherent one and two
qubit gating processes were controlled by time dependent localised classical
EM fields, the two qubit gating processes being facilitated by a high Q
cavity mode. The qubits were coupled to reservoir of spontaneous emission
(SE)\ modes, the cavity mode was coupled to a bath of cavity decay modes.
For ionic qubits, the numerous collective vibrational qubit modes also acted
as a reservoir, with Lamb-Dicke coupling to the internal qubit system. A key
objective of the work was to investigate decoherence effects due to the
qubit vibrational motion. Parameters similar to those in the model treated
by Tregenna et al \cite{Tregenna02a} were chosen, with comparable cavity
decay and cavity Rabi frequencies, both much larger than the spontaneous
emission decay rate and the Rabi frequencies of the two qubit gating fields.
One optical transition was resonant with the cavity mode. For two qubit
gating, the other transition was also assumed not coupled to the cavity
mode. Our primary aim was to evaluate fundamental rather than technical
causes of decoherence in standard qubit based quantum computers.\medskip

For the standard model we investigated, cavity decay, spontaneous emission
and Lamb-Dicke coupling to the vibrational modes were the most important
fundamental causes of decoherence. Effects due to Rontgen and diamagnetic
interactions were found to be negligible. Technical causes of decoherence,
such as fluctuations in the trapping fields, though needed to relate
decoherence times to current experiments \cite{Schmidt-Kaler03a} were not
studied here.\medskip

Characteristic decoherence time scales were evaluated for specific qubit
states (Hadamard, GHZ) at finite temperature in the situation with no gating
processes occurring. The decoherence time scaled as $1/N$. The decoherence
time scale for the uncorrelated Hadamard state could be made infinite by
choosing two optical dipole matrix elements that added to zero. The
decoherence time scale for the GHZ state was very long, due to the Boltzmann
factor - $\tau _{D}$\ being about 10$^{19}$s for $N\thickapprox $10$^{4}$\
qubits, even if the free SE decay rate of 10$^{8}$s$^{{\small -1}}$%
applied.\medskip

For the case of one qubit gating processes due to weak two photon resonant
Raman fields with a large one photon detuning, the decoherence time scale
was evaluated, but at zero temperature. Decoherence was mainly due to
Lamb-Dicke coupling of the gated qubit with the Raman fields, but the loss
of fidelity during the gating process was small, being proportional to the
square of the Lamb-Dicke parameter. Scaling effects were associated with
non-gated qubits and were small. For both optical transitions coupled to the
cavity mode, decoherence was associated with Lamb-Dicke coupling of
non-gated qubits with the cavity mode, no photon being present. However, the
effects were negligible even for $N\thickapprox $10$^{4}$\ qubits. Scaling
effects were absent for only one coupled transition.\medskip

For the case of two qubit gating processes due to one photon resonant gating
fields, as in the work of Tregenna et al \cite{Tregenna02a}, the decoherence
time scale was evaluated, also at zero temperature. Scaling effects were
absent for the parameters chosen, so overall the decoherence time is
independent of qubit numbers. However, other parameter choices, such as
having both optical transitions coupled to the cavity mode would lead to a
significant contribution associated with Lamb-Dicke coupling with the cavity
mode, one photon being present. In this case, modest qubit numbers $%
N\thickapprox $10$^{3}$\ qubits would result in non-gated contributions that
exceed those for the gated qubits. \medskip

In our model the parameters used have been the same as for the theoretical
model studied by Tregenna et al \cite{Tregenna02a}, rather than those where
real ions are involved. This was done in order to compare for quantum
computer models of this type, the effects of including (or not including)
the scaling up of qubit numbers and allowing for decoherence due to
vibrational motion. A treatment for real ions based on three \textit{state}
lambda systems and involving only \textit{electric dipole} transitions is
generally too simplified. The presence of other states (such as additional
magnetic substates, or states associated with other hyperfine levels) may
need to be taken into account, the actual transitions involved may be of
electric quadrupole or magnetic dipole character, and magnetic fields may
need to be present in order that only the 0-2 and 1-2 transitions are
resonant with the two-qubit gating laser fields. There are several
possibilities which involve storing the qubit in states 0, 1 and utilising
an excited state 2 in the gating processes, so that although these key
states form a lambda system, other states or non electric dipole transitions
may be involved. Consider the case where states 0, 1 are associated with two 
\textit{hyperfine} sublevels of a ground electronic level and state 2 is an
optical excited state. A simple system of this type involves a $^{2}$S$%
_{1/2} $ ground level and a $^{2}$P$_{1/2}^{o}$ excited state, but with a
non-zero nuclear spin $I=1/2$. With $^{2}$S$_{1/2}$ ($F=0,M_{F}=0$) as state
0, $^{2}$S$_{1/2}$ ($F=1,M_{F}=+1$) as state 1 and with $^{2}$P$_{1/2}^{o}$ (%
$F=1,M_{F}=+1$) as state 2, suitable polarisations for the gating laser
fields can be chosen to only cause transitions between these states.
However, spontaneous emission causes transitions into the $^{2}$S$_{1/2}$ ($%
F=1,M_{F}=0$) state, so there is no longer a three state lambda system. A
second example is where states 0, 1 are associated with two \textit{Zeeman}
substates of a ground electronic level and state 2 is an optical excited
state.A simple system of this type exists in $^{40}$Ca$^{+}$where states 0,
1 are the ground level $^{2}$S$_{1/2}$ ($M_{J}=-1/2$) and ($M_{J}=+1/2$)
states and state 2 is say the metastable $^{2}$D$_{5/2}$ ($M_{J}=+1/2$)
state. Here the nuclear spin is $I=0$, so no hyperfine structure is
involved. With suitable polarisations the required one and two qubit gating
processes that do not involve other states can be performed, the presence of
a non-zero magnetic field detuning the transition between $^{2}$S$_{1/2}$ ($%
M_{J}=-1/2$) and the additional $^{2}$D$_{5/2}$ ($M_{J}=-1/2$) state.
Spontaneous emission from state 2 only causes transitions to states 0, 1, so
here a genuine lambda system is involved. However, an electric quadrupole
transition connects state 2 with 0 and 1 rather than an electric dipole
transition. If the state 2 was chosen as say the lowest $^{2}$P$_{3/2}^{o}$ (%
$M_{J}=+1/2$) state so that electric dipole transitions are involved, then
spontaneous emission processes to $^{2}$D$_{5/2}$ and $^{2}$D$_{3/2}$ states
occur and more than three states would be involved. A final example involves
storing the qubit in states 0, 1 where 0 is a ground state and state 1 is a 
\textit{metastable excited} state, so that $\omega _{10}$ is an optical
rather than a Zeeman or hyperfine frequency. Such a case exists in $^{40}$Ca$%
^{+}$ with the states 0, 1 and 2 being magnetic substates of the lowest $%
^{2} $S$_{1/2}$, $^{2}$D$_{5/2}$ and $^{2}$P$_{3/2}^{o}$ energy levels (with
respective substates $M_{J}=1/2,5/2,3/2$ for example). However, even with
suitable laser field polarisations for the one and two qubit gating fields,
the additional $M_{J}=1/2,3/2$ substates of the $^{2}$D$_{5/2}$ level become
involved due to spontaneous emission from the $^{2}$P$_{3/2}^{o}$ ($%
M_{J}=3/2 $) state. Thus, the theory would need to be extended to allow for
the actual states and radiative transitions involved for a particular ion of
interest. This choice of ion would be made to minimise the numbers of states
needed - suggesting avoidance of cases where there are many lower energy
(fine structure, hyperfine stucture) levels, together with as small an upper
state spontaneous decay rate as possible - suggesting avoidance of electric
dipole downward transitions in favour of electric quadrupole or magnetic
dipole processes. The 0-2, 1-2 transitions also need to be in the optical
frequency range in order to couple these transitions to a high Q cavity.
Also, cases where there are other levels between 0,1 and 2 are also
unfavourable, as other such states may be populated via spontaneous
emission.\medskip

In conclusion, lambda systems localised in a high Q cavity, which can both
facilitate two qubit gating processes and reduce decoherence caused by
spontaneous emission and cavity decay, are a useful system for research on
scalable quantum computers. However, for real ions the model needs to be
expanded to include the presence of all magnetic substates and to treat the
case of electric quadrupole and magnetic dipole transitions. The case of
neutral qubits, where the vibrational modes are independent and do not
constitute a reservoir, is also of significant interest and a treatment via
the present Markovian theory would be worthwhile.\bigskip

\section{Acknowledgements{\protect\small \protect\medskip }}

The author is grateful for helpful discussions with A. Beige, S. Barnett, J.
Compagno, H. Carmichael, I. Deutsch, J. Eberly, J. Eschner, F. Haake, E.
Hinds, D. Jaksch, P. Knight, J. Pachos, D. Pegg, W. Phillips, M. Plenio, S.
Scheel, F. Schmidt-Kaler, B. Varcoe and H. Wiseman on various aspects of
this work. Helpful comments from a referee are also acknowledged.\bigskip

\section{Figure caption}

\textbf{Figure 1}. Model of an $N$ qubit quantum computer. Three-state
lambda system qubits are localised around well-separated positions via
trapping potentials, and undergo collective centre of mass (CM) vibrational
motions. Coherent one and two qubit gating processes are controlled by time
dependent localised classical electromagnetic (EM) fields that address
specific qubits. Two qubit gating processes are facilitated by a cavity mode
ancilla, which permits state interchange between qubits. The lambda system
qubits are coupled to a bath of EM field spontaneous emission (SE) modes,
and the cavity mode is coupled to a bath of cavity decay modes. For large $N$
the numerous collective vibrational modes of the qubits also act as a
reservoir, coupled to the qubits, the cavity mode and the SE modes.


\bigskip

\begin{thebibliography}{99}
\bibitem{Feynman80a} FEYNMAN, R. P., \textit{Int. J. Theor. Phys.}, 21, 467
(1982)\ 

\bibitem{Deutsch85a} DEUTSCH, D., \textit{Proc. Roy. Soc. Lond. A: Math.
Phys. Sci.}, 400, 97 (1985)\ 

\bibitem{Nielsen00a} NIELSEN, M. A. \& CHUANG, I. L., \textit{Quantum
Computation and Quantum Information}, Cambridge University Press (2000).

\bibitem{Shor94a} SHOR, P. W., \textit{Proc. 35th Annual Symposium on the
Foundations of Computer Science},.IEEE Computer Society Press\ (1994).
Eprint Quant-Ph/9508027.

\bibitem{Grover97a} GROVER, L., \textit{Phys. Rev. Letts.}, 79, 325 (1997).

\bibitem{Fahri01a} FAHRI, E., GOLDSTONE, J., GUTMAN, S., LAPAN, J.,
LUNDGREN, A. \& PREDA, D., \textit{Science}, 292, 472 (2001).

\bibitem{Gottesman01a} GOTTESMAN, D., KITAEV, A. \& PRESKILL, J., \textit{%
Phys. Rev. A}, 64, 012310 (2001).

\bibitem{Freedman03a} FREEDMAN, M. H., KITAEV, A., LARSEN, M. J. \& WANG,
Z., \textit{Bull. Amer. Math. Soc.}, 40, 31 (2003). Eprint Quant-Ph/0101025.

\bibitem{Duan01a} DUAN, L-M., CIRAC, J. I. \& ZOLLER, P., \textit{Science},
292, 1695 (2001).

\bibitem{Knill01a} KNILL, E., LAFLAMME, R. \& MILBURN, G. J., \textit{Nature}%
, 409, 46 (2001).

\bibitem{Nielsen99a} NIELSEN, M. A., \textit{Phys. Rev. Letts}., 86, 5188
(2001).

\bibitem{Raussendorf01a} RAUSSENDORF, R. \& BRIEGEL, H. J., \textit{Phys.
Rev. Letts.}, 83, 436 (1999).

\bibitem{Gulde03a} GULDE, S., RIEBE, M., LANCASTER, G., BECHER, C., ESCHNER,
J., HAFFNER, R., SCHMIDT-KALER, F., CHUANG, I. \& BLATT, R., \textit{Nature}%
, 421, 48 (2003).

\bibitem{Vandersypen01a} VANDERSYPEN, L., STEFFEN, M., BREYTA, G., YANNONI,
C., SHERWOOD, M. \& CHUANG, I., \textit{Nature}, 414, 883 (2001).

\bibitem{Cirac95a} CIRAC, J. I. \& ZOLLER, P., \textit{Phys. Rev. Letts.},
74, 4091 (1995).

\bibitem{Kielpinski02a} KIELPINSKI, D., MONROE, C. \& WINELAND, D. J., 
\textit{Nature}, 417, 709 (2002).

\bibitem{Monroe02a} MONROE, C., \textit{Nature}, 416, 238 (2002).

\bibitem{Cirac04a} CIRAC, J. I., DUAN, L-M. \& ZOLLER, P., (2004). \textit{%
Eprint Quant-Ph}/0405030.

\bibitem{DiVincenzo00a} DIVINCENZO, D., \textit{Fortsch. fur Phys.}, 48, 771
(2000).

\bibitem{LANL04a} HUGHES, R. \& HEINRICHS, T., Eds., \textit{A Quantum
Information Science and Technology Roadmap}, http://qist.lanl.gov (2004).

\bibitem{Fedichkin03a} FEDICHKIN, L., FEDOROV, A. \& PRIVMAN, V., \textit{%
Proceedings of SPIE}, 5105, 243 (2003). Eds. E. Donkor, A. R. Pirich \& H.
E. Brandt.

\bibitem{Zurek03a} ZUREK, W. H., \textit{Rev. Mod. Phys.}, 75, 715 (2003).

\bibitem{Guilini96a} GUILINI, D., JOOS, E., KIEFER, C., KUPSCH, J.,
STAMATESCU, I. O. \& ZEH, H. D., \textit{Decoherence and the Appearance of a
Classical World}, Springer-Verlag (1996).

\bibitem{Shor95a} SHOR, P., \textit{Phys. Rev. A}, 52, R2493 (1995).

\bibitem{Calderbank96a} CALDERBANK, A. R. \& SHOR, P., \textit{Phys. Rev. A}%
, 54, 1098 (1996).

\bibitem{Steane96a} STEANE, A., \textit{Phys. Rev. Letts.}, 77, 793 (1996).

\bibitem{Viola98a} VIOLA, L. \& LLOYD, S., \textit{Phys. Rev. A}, 58, 2733
(1998).

\bibitem{Zanardi97a} ZANARDI, P. \& RASETTI, M., \textit{Phys. Rev. Letts}.,
79, 3306 (1997).

\bibitem{Lidar98a} LIDAR, D. A., CHUANG, I. L. \& WHALEY, K. B., \textit{%
Phys. Rev. Letts.}, 81, 2594 (1998).

\bibitem{Duan98a} DUAN, L-M. \& GUO, G-C., \textit{Phys. Rev. A}, 58, 3491
(1998).

\bibitem{Beige00a} BEIGE, A., BRAUN, D., TREGENNA, B. \& KNIGHT, P. L., 
\textit{Phys. Rev. Letts.}, 85, 1762 (2000).

\bibitem{Byrd03a} BYRD, M. S. \& LIDAR, D. A., \textit{J. Mod. Opt.}, 50,
1285 (2003).

\bibitem{Preskill99a} PRESKILL, J., \textit{Physics Today}, June 1999, p24.

\bibitem{Preskill98a} PRESKILL, J., \textit{Proc. Roy. Soc. Lond. A: Math.
Phys. Sci.}, 454, 385 (1998).

\bibitem{Knill98a} KNILL, E., LAFLAMME, R. \& ZUREK, W. H., \textit{Proc.
Roy. Soc. Lond. A: Math. Phys. Sci.}, 454, 365 (1998).

\bibitem{Namiki03a} NAMIKI, M., PASCAZIO, S. \& NAKAZATO, H., \textit{%
Decoherence and Quantum Measurements}, World Scientific (2003).

\bibitem{Dalton04a} DALTON, B. J., \textit{Unpublished work} (2004).

\bibitem{Tolkunov04a} TOLKUNOV, D. \& PRIVMAN, V., \textit{Phys. Rev. A},
69, 062309 (2004).

\bibitem{Braun01a} BRAUN, D., HAAKE, F. \& STRUNZ, W.H., \textit{Phys. Rev.
Letts.}, 86, 2913 (2001).

\bibitem{Dalton03a} DALTON, B. J., \textit{J. Mod. Opt.}, 50, 951 (2003).

\bibitem{Plenio96a} PLENIO, M. B. \& KNIGHT, P. L., \textit{Phys. Rev. A},
53, 2986 (1996).

\bibitem{Plenio97a} PLENIO, M. B. \& KNIGHT, P. L., \textit{Proc. Roy. Soc.
Lond. A: Math. Phys. Sci.}, 453, 2017 (1997)\ 

\bibitem{Cirac00a} CIRAC, J. I. \& ZOLLER, P., \textit{Nature}, 406, 579
(2000).

\bibitem{Calarco01a} CALARCO, T., CIRAC, J. I. \& ZOLLER, P., \textit{Phys.
Rev. A}, 63, 062304 (2001).

\bibitem{DeVoe98a} DEVOE, R. G., \textit{Phys. Rev. A}, 58, 910 (1998).

\bibitem{Duan04a} DUAN, L-M., \textit{Phys. Rev. Letts.}, 93, 100502 (2004).

\bibitem{Pellizzari95a} PELLIZZARI, T., GARDINER, S. A., CIRAC, J. I. \&
ZOLLER, P., \textit{Phys. Rev. Letts.}, 75, 3788 (1995).

\bibitem{Domokos95a} DOMOKOS, P., RAIMOND, J. M., BRUNE, M. \& HAROCHE, S., 
\textit{Phys. Rev. A}, 52, 3554 (1995).

\bibitem{Turchette95a} TURCHETTE, Q. A., HOOD, C. J., LANGE, W., MABUCHI, H.
\& KIMBLE, H. J., \textit{Phys. Rev. Letts.}, 75, 4710 (1995).

\bibitem{Sauer04a} SAUER, J. A., FORTIER, K. M., CHANG, M. S., HAMLEY, C. D.
\& CHAPMAN, M. S., \textit{Phys. Rev. A}, 69, 051804 (2004).

\bibitem{Duan04b} DUAN, L-M., BLINOV, B. B., MOEHRING, D. L. \& MONROE, C.,
(2004). \textit{Eprint Quant-Ph}/0401020.

\bibitem{Guthohrlein01a} GUTHOHRLEIN, G. R., KELLER, M., HAYASAKA, K.,
LANGE, W.. \& WALTHER, H. , \textit{Nature}, 414, 49 (2001).

\bibitem{Mundt02a} MUNDT, A. B., KREUTER, A., BECHER, C., LEIBFRIED, D.,
ESCHNER, J., SCHMIDT-KALER, F. \& BLATT, R., \textit{Phys. Rev. Letts.}, 89,
103001 (2002).

\bibitem{Beige00b} BEIGE, A., BOSE, S., BRAUN, D., HUELGA, S. F., KNIGHT, P.
L., PLENIO, M. B. \& VEDRAL, V., \textit{J. Mod. Opt.}, 47, 2583 (2000).

\bibitem{Pachos02a} PACHOS, J. \& WALTHER, H., \textit{Phys. Rev. Letts.},
89, 187903 (2002).

\bibitem{Tregenna02a} TREGENNA, B., BEIGE, A. \& KNIGHT, P.L., \textit{Phys.
Rev. A}, 65, 032305 (2002).

\bibitem{Beige04a} BEIGE, A., CABLE, H., MARR, C. \& KNIGHT, P. L., (2004). 
\textit{Eprint Quant-Ph}/0405186.

\bibitem{Garg96a} GARG, A., \textit{Phys. Rev. Letts.}, 77, 964 (1996).

\bibitem{Schmidt-Kaler03a} SCHMIDT-KALER, F., GULDE, S., RIEBE, M.,
DEUSCHLE, T., KREUTER, A., LANCASTER, G., BECHER, C., HAFFNER, R. \& BLATT,
R., \textit{J. Phys. B:} \textit{Atom. Mol. Phys.}, 36, 623 (2003).

\bibitem{Barnett96a} BARNETT, S. M., BURNETT, K. \& VACCARO, J. A., \textit{%
J. Res. Nat. Inst. Stand. \& Tech.}, 101, 593 (1996).

\bibitem{Vitanov97a} VITANOV, N.V. \& STENHOLM, S., \textit{Phys. Rev. A},
55, 648 (1997).

\bibitem{Ziman65a} ZIMAN, J. M., \textit{Principles of the Theory of Solids}%
, Cambridge University Press (1965).\newpage
\end{thebibliography}
\end{document}